\documentclass[final,3p,times,fleqn]{elsarticle}

\usepackage{amssymb}
\usepackage{mathrsfs}
\usepackage{graphicx}
\usepackage{bm}

\newcommand{\Sp}{\mathrm{Sp}}

\renewcommand{\H}{{\mathcal H}}
\newcommand{\irm}{\mathrm{i}}

\renewcommand{\L}{{\mathcal L}}

\journal{arXiv.org}

\begin{document}

\begin{frontmatter}

\title{Prototype and reduced nonlinear integrable lattice systems with the  modulated pulson   behavior}

\author{Oleksiy O.~Vakhnenko}

\address{Department for Theory of Nonlinear Processes in Condensed Matter,
Bogolyubov  Institute for Theoretical Physics,
The National Academy of Sciences of Ukraine,
14-B Metrologichna Street, Ky\"{\i}v 03143, Ukra\"{\i}na}\ead{vakhnenko@bitp.kiev.ua}

\begin{abstract}
A multi-component semi-discrete nonlinear integrable system  associated with the relevant third-order auxiliary linear problem is claimed to be the  prototype system for several reduced integrable systems formulated in terms of true dynamical field variables. The main conservation laws related to the prototype system are found in the framework of generalized recurrent approach. The two-fold Darboux-B\"{a}cklund dressing technique  as applied to the integration of prototype system is developed in details. The novel reduced complex-valued  nonlinear integrable system embracing three coupled dynamical subsystems on a quasi-one-dimensional lattice is proposed and its concise Lagrangian and Hamiltonian representations are  written down. The essentially nontrivial couplings between the two complex-valued Toda-like subsystems are shown to be mediated by the intermediate subsystem both in their kinetic and potential parts. The explicit multi-component solution of a modulated pulson type related to the reduced integrable nonlinear  ternary system is presented and analyzed.
\end{abstract}


\end{frontmatter}

\section{Introduction}
\label{sec1}
\setcounter{equation}{0}

The seminal suggestion of one-dimensional dynamical lattice system with the  exponential nonlinearity by Toda \cite{toda-JPSJ-22-431, toda-JPSJ-23-501, toda-Butsuri-51-185} and the subsequent discovery of its complete integrability by Manakov \cite{manakov-ZETF-67-543} and  Flaschka \cite{flaschka-PTP-51-703} had undoubtedly been and continue to be the powerful motivating forces for the  widespread development of numerous   semi-discrete (\textit{i.e.} continuous in time and discrete in spatial coordinate)  integrable models \cite{ablowitz-16-598, ablowitz-JMP-17-1011, hirota-JPSJ-35-289, hirota-JPSJ-40-891, inozemtsev-CMP-121-629, suris-JPA-30-2235,  vakhnenko-JMP-51-103518, bogdan-JPSJ-83-064007} including the  multi-component ones \cite{bruschi-JMP-21-2749, gerdzhikov-TMF-52-89, tsuchida-JPA-32-2239, tsuchida-JPA-35-7827, ablowitz-PLA-253-287, vakhnenko-JPA-32-5735}. This stream of researches  is   accompanied  by the recognized applicability of semi-discrete integrable nonlinear systems to  the various branches of physics \cite{vakhnenko-JMP-56-033505} inasmuch as the vast number of real physical objects are given on one or another type of quasi-one-dimensional lattices.

Among other  semi-discrete nonlinear integrable systems   the integrable systems generated by the various auxiliary  linear problems of third \cite{levi-JPA-35-L67, vakhnenko-JPA-36-5405, vakhnenko-UJP-48-653, vakhnenko-JNMP-20-606, vakhnenko-CSF-60-1, vakhnenko-WM-88-1, vakhnenko-PLA-384-126081} or more higher \cite{vakhnenko-JNMP-18-401,  vakhnenko-JNMP-18-415}  order deserve a special attention as the systems comprising several subsystems of distinct physical origins.

Thus, in two our papers \cite{vakhnenko-JPA-36-5405, vakhnenko-UJP-48-653} we have introduced the general integrable nonlinear system
\begin{equation}\label{eq1.1}
\dot{p}_{11}(n)=F_{12}(n)G_{21}(n-1)-F_{12}(n+1)G_{21}(n)
\end{equation}
\begin{equation}\label{eq1.2}
\dot{p}_{13}(n)=F_{12}(n)G_{23}(n-1)-F_{12}(n+1)G_{23}(n)
\end{equation}
\begin{equation}\label{eq1.3}
\dot{p}_{31}(n)=F_{32}(n)G_{21}(n-1)-F_{32}(n+1)G_{21}(n)
\end{equation}
\begin{equation}\label{eq1.4}
\dot{p}_{33}(n)=F_{32}(n)G_{23}(n-1)-F_{32}(n+1)G_{23}(n)
\end{equation}
\begin{equation}\label{eq1.5}
\dot{F}_{12}(n)=p_{11}(n)F_{12}(n)+p_{13}(n)F_{32}(n)
\end{equation}
\begin{equation}\label{eq1.6}
\dot{G}_{21}(n)=-G_{21}(n)p_{11}(n)-G_{23}(n)p_{31}(n)
\end{equation}
\begin{equation}\label{eq1.7}
\dot{G}_{23}(n)=-G_{21}(n)p_{13}(n)-G_{23}(n)p_{33}(n)
\end{equation}
\begin{equation}\label{eq1.8}
\dot{F}_{32}(n)=p_{31}(n)F_{12}(n)+p_{33}(n)F_{32}(n)
\end{equation}
permitting a number of reductions from the so-called prototype $p_{11}(n)$, $p_{13}(n)$, $p_{31}(n)$, $p_{33}(n)$, $F_{12}(n)$, $G_{21}(n)$, $G_{23}(n)$, $F_{32}(n)$
to the actual physical field variables.
Here the overdot  marks  the derivative with respect to time  $\tau$, while  the discrete spatial coordinate $n$ is assumed to span all integers from minus to plus infinity.  To integrate this system (\ref{eq1.1})--(\ref{eq1.8})  we have developed rather complicated inverse scattering technique and found some simplest  solutions related to the reduced system encompassing two real-valued Toda-like translational subsystems coupled with one orientational subsystem. Unfortunately, the orientational component of the obtained solution suffers to be unexcited \cite{vakhnenko-JPA-36-5405, vakhnenko-UJP-48-653}.

In present paper we try to fill in this gap   in the framework of specially developed  two-fold Darboux--Backlund dressing approach which allow to construct  the nonlinear wave packet embracing all involved field components. In addition, we apply the general recurrent scheme  capable to generate the infinite hierarchy  of local conservation laws and write down  the most important  of them explicitly.  Moreover, we propose early unknown reduction to  the actual field variables appropriate for the practical implementation of obtained results to the low-dimensional regular physical objects. We formulate the   reduced nonlinear  integrable system both  in terms of Lagrange and Hamilton dynamic approaches and underline that two of its three coupled subsystems have  a nontrivial fashion of complex-valued Toda-like type. In so doing,  the  respective Hamilton function is proved to emanate from one of the conserved quantities.

To proceed with the most coherent consideration concerning the  general nonlinear system (\ref{eq1.1})--(\ref{eq1.8}) we  begin with the  confirmation of its complete integrability.

\section{Complete integrability of the general nonlinear system}
\label{sec2}
\setcounter{equation}{0}

One can readily verify that the general semi-discrete nonlinear system of our interest (\ref{eq1.1})--(\ref{eq1.8}) is presentable in the form of matrix-valued semi-discrete zero-curvature equation
\begin{equation}\label{eq2.1}
\dot{L}(n|z)=A(n+1|z)L(n|z)-L(n|z)A(n|z)
\end{equation}
with the spectral $L(n|z)$ and evolution $A(n|z)$ operators given by the following $3\times3$ square matrices
\begin{equation}\label{eq2.2}
L(n|z) = \left( \begin{array}{ccccc}
p_{11}(n)+\lambda(z) &&  F_{12}(n) && p_{13}(n)\\
G_{21}(n) && 0 && G_{23}(n)\\
p_{31}(n ) && F_{32}(n) && p_{33}(n)+\lambda(z)\\
\end{array}\right)
\end{equation}
\begin{equation}\label{eq2.3}
A(n|z) =  \left( \begin{array}{ccccc}
0 &&  -F_{12}(n) && 0\\
-G_{21}(n-1) && \lambda(z) && -G_{23}(n-1)\\
0 && -F_{32}(n) && 0\\
\end{array} \right)\,.
\end{equation}
Here the symbols $\lambda(z)$ and $z$ denote the authentic and rationalizes time-independent spectral parameters, respectively. According to the commonly referred terminology \cite{newell-SMP, takhtadzhyan-GPTS, tu-JPA-22-2375} this representation (\ref{eq2.1})--(\ref{eq2.3}) determines the complete integrability of the declared semi-discrete nonlinear  system (\ref{eq1.1})--(\ref{eq1.8}) in the Lax sense.

Let us remind that the zero-curvature equation itself (\ref{eq2.1}) serves as the compatibility condition between two  linear matrix-valued equations
\begin{equation}\label{eq2.4}
X(n+1|z)=L(n|z)X(n|z)
\end{equation}
\begin{equation}\label{eq2.5} 
\dot{X}(n|z)=A(n|z)X(n|z)
\end{equation}
referred to as the auxiliary linear problem. Here the notation $X(n|z)$ stands for the auxiliary matrix-function of discrete spatial coordinate $n$, continuous time $\tau$ and time-independent spectral parameter $z$. By and large, the auxiliary linear problem (\ref{eq2.4}) and (\ref{eq2.5}) is capable to lay the foundation for the development of various methods of system's  integration such as the  method of inverse scattering transform \cite{caudrey-NHMS-97-221, vakhnenko-JMP-51-103518, vakhnenko-JPA-36-5405, vakhnenko-UJP-48-653} and the Darboux-B\"{a}cklund dressing method  \cite{matveev-LMP-3-217, matveev-LMP-3-425, chowdhury-LMP-7-313, vakhnenko-JPSJ-84-014003,  vakhnenko-EPJP-133-243}. Moreover, it might provide an  indispensable starting tool in  recurrent searching for  an infinite hierarchy of local conservation laws \cite{vakhnenko-JNMP-18-401}.

\section{Natural constraints and the main local  densities}
\label{sec3}
\setcounter{equation}{0}

It is well known that any  nonlinear integrable  system on an infinite regular chain possesses the infinite number of  local conservation laws.
The most straightforward way to isolate some of them is based upon the universal local conservation law
\begin{equation}\label{eq3.1}
\frac{d}{d\tau}\ln [\det L(n|z)]=\Sp A(n+1|z) -\Sp A(n|z)
\end{equation}
appearing as a simple contraction of  system's zero-curvature representation (\ref{eq2.1}).

Inasmuch as the determinant $\det L(n|z)$ of spectral matrix $L(n|z)$  depends on two distinct powers of spectral parameter $\lambda(z)$ and the expression $\Sp A(n+1|z) -\Sp A(n|z)$ is equal to zero,  the universal local conservation law (\ref{eq3.1}) produces two unicellular   conservation laws imposing two natural  constraints onto the set of prototype field variables.   The explicit record of natural constraints depends on a particular choice of boundary conditions for the prototype field variables and on an  expected physical sense  of reduced field variables. Assuming the  underlying  lattice being spatially  uniform and  pinned to  the immovable frame of reference we take  the natural constraints in the following form
\begin{equation}\label{eq3.2}
p_{11}(n)F_{32}(n)G_{23}(n)-p_{31}(n)F_{12}(n)G_{23}(n)+
p_{33}(n)F_{12}(n)G_{21}(n)-p_{13}(n)F_{32}(n)G_{21}(n)=0
\end{equation}
\begin{equation}\label{eq3.3}
F_{12}(n)G_{21}(n)+F_{32}(n)G_{23}(n)=-1\,.
\end{equation}

These  constraints (\ref{eq3.2}) and (\ref{eq3.3}) are consistent with the following boundary conditions for the prototype fields:
$\lim_{n\rightarrow-\infty} p_{11}(n)\rightarrow0$, $\lim_{n\rightarrow-\infty} p_{13}(n)\rightarrow0$, $\lim_{n\rightarrow-\infty} p_{33}(n)\rightarrow0$, $\lim_{n\rightarrow-\infty} p_{31}(n)\rightarrow0$ and
$\lim_{n\rightarrow-\infty} F_{12}(n)\rightarrow F_{12}$, $\lim_{n\rightarrow-\infty} G_{21}(n)\rightarrow G_{21}$, $\lim_{n\rightarrow-\infty} F_{32}(n)\rightarrow F_{32}$,
$\lim_{n\rightarrow-\infty} G_{23}(n)\rightarrow G_{23}$,   where $F_{12}G_{21}+F_{32}G_{23}=-1$.
As a consequence, the limiting eigenvalue problem
\begin{equation}\label{eq3.4}
L(z)X(z)=X(z)\zeta(z)
\end{equation}
(where $L(z)=\lim_{n\rightarrow-\infty}L(n|z)$)  prescribes the functional relationship $\lambda(z)=z+1/z$  in view of very simple resulting expressions
\begin{equation}\label{eq3.5}
\zeta_{1}(z)=z
\end{equation}
\begin{equation}\label{eq3.6}
\zeta_{2}(z)=\lambda(z)\equiv z+1/z
\end{equation}
\begin{equation}\label{eq3.7}
\zeta_{3}(z)=1/z
\end{equation}
for the eigenvalues $\zeta_{j}(z)$. It is interesting to note that the functional dependence $\lambda(z)=z+1/z$ establishes a particular realization of the   Zhukovskiy transformation \cite{joukowsky-ZFM-1-281, joukowsky-ZFM-3-81} well known in aerodynamics.

The  capability  of   universal local conservation law (\ref{eq3.1}) in generating system's local conservation laws is seen to be    restricted only by two specimens.

In contrast, there exists  the generalized direct procedure \cite{vakhnenko-JNMP-18-401, vakhnenko-JNMP-20-606,vakhnenko-WM-88-1, vakhnenko-PLA-384-126081} permitting to develop an infinite set of local conservation laws recursively without   references to auxiliary spectral data. By definition, any local conservation law associated with some semi-discrete   system given on an infinite quasi-one-dimensional lattice can be written in the form
\begin{equation}\label{eq3.8}
\dot{\rho}(n) = J(n|n-1)-J(n+1|n)\,,
\end{equation}
where the quantities $\rho(n)$ and $J(n+1/2|n-1/2)$ are referred to as  the local density and the local current, respectively. Bearing in mind this general definition (\ref{eq3.8}) we must find the  recursive  presentation (\textit{i.e}  presentation  based upon some proper expansion in spectral parameter $z$ or inverse spectral parameter $1/z$)
for the auxiliary quantities $\Gamma_{jk}(n|z)$ governed by the following set of spatial Riccati equations
\begin{equation}\label{eq3.9}
\Gamma_{jk}(n+1|z) \sum_{i=1}^{3} L_{ki}(n|z) \Gamma_{ik}(n|z) =
\sum_{i=1}^{3} L_{ji}(n|z) \Gamma_{ik}(n|z)
\end{equation}
with  the restrictions
\begin{equation}\label{eq3.10}
\Gamma_{ji}(n|z)\Gamma_{ik}(n|z)=\Gamma_{jk}(n|z)
\end{equation}
being taken into account. The obtained series should be substituted  into the collection of three ($j=1, 2, 3$) generating equations
\begin{equation}\label{eq3.11}
\frac{d}{d\tau}\ln M_{jj}(n|z) =B_{jj}(n+1|z) -B_{jj}(n|z)\,.
\end{equation}
Here the composite functions
\begin{equation}\label{eq3.12}
M_{jj}(n|z) =\sum_{i=1}^{3} L_{ji}(n|z) \Gamma_{ij}(n|z)
\end{equation}
and
\begin{equation}\label{eq3.13}
B_{jj}(n|z) =\sum_{i=1}^{3} A_{ji}(n|z) \Gamma_{ij}(n|z)
\end{equation}
serve to  generate the hierarchy of local densities and the hierarchy of local currents, respectively. In so doing, the quantities $L_{jk}(n|z)$ and $A_{jk}(n|z)$ denote the matrix elements of spectral  $L(n|z)$ and evolution $A(n|z)$ operators, respectively. Collecting terms with the same powers of spectral parameter in each of three ($j=1, 2, 3$) generating series (\ref{eq3.11}) it is possible to recover any required number  of local conservation laws  from their infinite hierarchy.

The most productive is the second ($j=2$) of generating series (\ref{eq3.11}). To develop the second generating series it is sufficient to consider  two auxiliary functions  $\Gamma_{12}(n|z)$ and $\Gamma_{32}(n|z)$ since $\Gamma_{jj}(n|z)\equiv1$ in view of properties (\ref{eq3.10}). Due to the evident symmetry $\lambda(z)=\lambda(1/z)$ of the functional spectral  parameter $\lambda(z)$ we restrict ourselves  only to  serial expansions at the center $|z|\rightarrow0$ of a rationalized complex spectral  plane   and seek the auxiliary functions  $\Gamma_{12}(n|z)$ and $\Gamma_{32}(n|z)$ as follows
\begin{equation}\label{eq3.14}
\Gamma_{12}(n|z) =z \sum_{j=0}^\infty \gamma_{12}(n|m)z^{m}
\end{equation}
\begin{equation}\label{eq3.15}
\Gamma_{32} (n|z) =z\sum_{j=0}^\infty \gamma_{32}(n|m)z^{m}~.
\end{equation}
Then for the   generating function $\ln M_{22}(n|z)$ written  up to the second power in spectral parameter $z$ we obtain
\begin{eqnarray}\label{eq3.16}
\ln M_{22}(n|z)&=&\ln z - \left[p_{11}(n)+p_{33}(n)\right]z+\nonumber\\
&+&\left[p_{11}^{2}(n)/2+p_{11}^{2}(n)/2+p_{13}(n)p_{31}(n)
-G_{21}(n)F_{12}(n+1)-G_{23}(n)F_{32}(n+1)-1\right]z^{2}\,.
\end{eqnarray}
By virtue of second generating equation (\textit{i. e.} equation (\ref{eq3.11}) taken at $j=2$) the quantities
\begin{equation}\label{eq3.17}
p(n)=p_{11}(n)+p_{33}(n)
\end{equation}
\begin{equation}\label{eq3.18}
h(n)=p_{11}^{2}(n)/2+p_{11}^{2}(n)/2+p_{13}(n)p_{31}(n)
-G_{21}(n)F_{12}(n+1)-G_{23}(n)F_{32}(n+1)-1
\end{equation}
acquire the status of local densities in two the most important local conservation laws.

The comprehensive analysis shows that the former (\ref{eq3.17}) of these local densities should be identified with the density of system's momentum, whereas the latter one (\ref{eq3.18}) -- with the density of system's energy.

By invoking  the revealed   natural constraints (\ref{eq3.2}) and (\ref{eq3.3}) it is possible  to replace  eight original  prototype field variables by means of six properly chosen actual dynamical field variables. A particular realization of  such a reduction is not unique.  However, in any case  we come to some   reduced  nonlinear lattice system comprising three coupled dynamical subsystems.   As to the   local density $h(n)$  its expression (\ref{eq3.18})  will be useful while writing   the Hamilton function of reduced  nonlinear dynamical system  in terms actual fields.

We postpone both of above tasks up to the  Section 9 and concentrate now on the development of   Darboux--B\"{a}cklund dressing integration scheme in terms of prototype field variables applicable to the  general  nonlinear system of our interest (\ref{eq1.1})--(\ref{eq1.8}).

\section{Fundamentals of  the  Darboux--B\"{a}cklund dressing technique  }
\label{sec4}
\setcounter{equation}{0}

In this Section we itemize the main steps elucidating   the essence   of the Darboux--B\"{a}cklund dressing scheme suitable for the integration of  the prototype nonlinear lattice system under study (\ref{eq1.1})--(\ref{eq1.8}).

We  start with the general  definition of Darboux transformation \cite{chowdhury-LMP-7-313}
\begin{equation}\label{eq4.1}
^{c}X(n|z)=~^{cs}D(n|z)~^{s}X(n|z)\,.
\end{equation}
The Darboux transformation (\ref{eq4.1}) connects the  seed (\textit{a priori} known) $^{s}X(n|z)$  and crop (required) $^{c}X(n|z)$  solutions of the auxiliary linear problem (\ref{eq2.4}) and (\ref{eq2.5}) via the Darboux matrix $^{cs}D(n|z)$, which should be  properly chosen in accordance with the particular realizations (\ref{eq2.2}) and (\ref{eq2.3}) of auxiliary  spectral $L(n|z)$ and evolution $A(n|z)$ operators. The Darboux matrix $^{cs}D(n|z)$    must obey  the set of matrix equations
\begin{equation}\label{eq4.2}
^{cs}D(n+1|z)~^{s}L(n|z)=~^{c}L(n|z)~^{cs}D(n|z)
\end{equation}
\begin{equation}\label{eq4.3}
^{cs}\dot{D}(n|z)=~^{c}A(n|z)~^{cs}D(n|z)-~^{cs}D(n|z)~^{s}A(n|z)
\end{equation}
serving as the spatial (\ref{eq4.2}) and temporal (\ref{eq4.3}) compatibility conditions between the seed
\begin{equation}\label{eq4.4}
^{s}X(n+1|z)=~^{s}L(n|z)~^{s}X(n|z)
\end{equation}
\begin{equation}\label{eq4.5}
^{s}\dot{X}(n|z)=~^{s}A(n|z)~^{s}X(n|z)
\end{equation}
and crop
\begin{equation}\label{eq4.6}
^{c}X(n+1|z)=~^{c}L(n|z)~^{c}X(n|z)
\end{equation}
\begin{equation}\label{eq4.7}
^{c}\dot{X}(n|z)=~^{c}A(n|z)~^{c}X(n|z)
\end{equation}
embodiments  of auxiliary linear problem (\ref{eq2.4}) and  (\ref{eq2.5}). Here and  later on the upper-front  single index $s$ or $c$ indicates  a quantity associated with the seed or crop solution, respectively.
The upper-front double index $cs$ marks a quantity related to or induced by the Darboux matrix.
According to the general rules  \cite{vakhnenko-JMP-56-033505, vakhnenko-JPSJ-84-014003, vakhnenko-EPJP-133-243}
the spatial compatibility condition (\ref{eq4.2})   establishes the implicit B\"{a}cklund transformation between the seed and crop   solutions for the prototype   fields, while the temporal  compatibility condition (\ref{eq4.3})
allows to uncover  the crucial spectral properties of  Darboux matrix sufficient to restore explicitly its components indispensable for the development of the whole   Darboux-B\"{a}cklund dressing integration scheme.

Seeking for the Darboux matrix $^{cs}D(n|z)$ we assume   the following two-fold ($c=s+2$) ansatz
\begin{equation}\label{eq4.8}
^{cs}D(n|z) = \left( \begin{array}{ccccc}
^{cs}K_{11}(n) &~&  ^{cs}C_{12}(n)\lambda(z) +~^{cs}T_{12}(n) &~& ^{cs}K_{13}(n)\\[2mm]
^{cs}E_{21}(n)\lambda(z) +~^{cs}V_{21}(n)&~& \lambda^{2}(z) +~^{cs}D_{22}(n)\lambda(z)+~^{cs}U_{22}(n) &~& ^{cs}E_{23}(n)\lambda(z) +~^{cs}V_{23}(n)\\[2mm]
^{cs}K_{31}(n ) &~& ^{cs}C_{32}(n)\lambda(z) +~^{cs}T_{32}(n) &~& ^{cs}K_{33}(n)\\
\end{array}\right),
\end{equation}
which proves to be  consistent with the set of governing matrix equations (\ref{eq4.2})--(\ref{eq4.3}).  The notion ``two-fold ($c=s+2$)'' implies that  each matrix element of adopted ansatz (\ref{eq4.8}) written  as a polynomial of the  spectral parameter $\lambda(z)$  has  the same leading term as the respective matrix element appearing in the product of two evolution matrices.
The  Darboux matrix taken in the two-fold form  (\ref{eq4.8}) are able to generate nontrivial spatially finite solutions (from the vacuum one)  embracing  all three subsystems of  a  reduced nonlinear dynamical system.

The spectral properties of Darboux matrix $^{cs}D(n|z)$ follow  directly from the contracted form
\begin{equation}\label{eq4.9}
\frac{d}{d\tau} \ln [\det\,^{cs}D(n|z)] =\Sp\,^{c}A(n|z) -\Sp\,^{s}A(n|z)
\end{equation}
of temporal compatibility condition (\ref{eq4.3}). Indeed, in view of the identity
\begin{equation}\label{eq4.10}
\Sp\,^{c}A(n|z) -\Sp\,^{s}A(n|z)\equiv0\,,
\end{equation}
dictated by the specific form  (\ref{eq2.2})  of evolution matrix $A(n|z)$,
the contracted equation (\ref{eq4.9}) with the adopted ansatz (\ref{eq4.8}) for the Darboux matrix $~^{cs}D(n|z)$ yields
\begin{equation}\label{eq4.11}
\det\,^{cs}D(n|z)=\left[\lambda(z)-\lambda(~^{cs}z_{+})\right]\left[\lambda(z)-\lambda(~^{cs}z_{-})\right]~^{cs}W(n)\,,
\end{equation}
where the quantity ~$^{cs}W(n)$ and the  spectral data ~$^{cs}z_{+}$, ~$^{cs}z_{-}$ are proved to be time-independent ~$^{cs}\dot{W}(n)=0$, ~$^{cs}\dot{z}_{+}=0$, ~$^{cs}\dot{z}_{-}=0$.

Evidently, $\det\,^{cs}D(n|~^{cs}z_{\pm})=0$ and hence  the Darboux transformation (\ref{eq4.1}) yields  $\det\,^{c}X(n|~^{cs}z_{\pm})=0$. Therefore  the condition
\begin{equation}\label{eq4.12}
\sum_{k=1}^{3} ~^{c}X_{jk}(n|\,^{cs}z_{\pm})~^{cs}\varepsilon_k(\,^{cs}z_{\pm})=0
\end{equation}
should be satisfied.  The detailed form of this condition (\ref{eq4.12}) is given by formula
\begin{equation}\label{eq4.13}
\sum_{i=1}^{3} \sum_{k=1}^{3} \,^{cs}D_{ji}(n|\,^{cs}z_{\pm}) ~^{s}X_{ik}(n|\,^{cs}z_{\pm})~^{cs}\varepsilon_k(\,^{cs}z_{\pm})=0\,.
\end{equation}
Here the functions ~$^{cs}D_{ji}(n|z)$ and ~$^{s}X_{ik}(n|z)$ stand for the elements of respective matrices ~$^{cs}D(n|z)$ and ~$^{s}X(n|z)$, while the time- and space-independent parameters ~$^{cs}\varepsilon_k(\,^{cs}z_{\pm}$) serve as the  spectral data. The invariability  of spectral parameters ~$^{cs}\varepsilon_k(\,^{cs}z_{\pm}$) in time and space  has a status of rigorously proved theorem.

The obtained condition (\ref{eq4.13}) encompasses six linear equations for fourteen Darboux functions $~^{cs}K_{11}(n)$,  $~^{cs}C_{12}(n)$, $~^{cs}T_{12}(n)$, $~^{cs}K_{13}(n)$, $~^{cs}E_{21}(n)$, $~^{cs}V_{21}(n)$, $~^{cs}D_{22}(n)$, $~^{cs}U_{22}(n)$, $~^{cs}E_{23}(n)$, $~^{cs}V_{23}(n)$, $~^{cs}K_{31}(n)$, $~^{cs}C_{32}(n)$, $~^{cs}T_{32}(n)$, $~^{cs}K_{33}(n)$. Fortunately only six  $~^{cs}C_{12}(n)$, $~^{cs}T_{12}(n)$, $~^{cs}D_{22}(n)$, $~^{cs}U_{22}(n)$, $~^{cs}C_{32}(n)$, $~^{cs}T_{32}(n)$  of them  turn out to be truly independent. We prove this statement in Section 5  by invoking the  implicit  B\"{a}cklund transformation (\ref{eq4.2}) rewritten  as twenty two detailed equations (\ref{eq5.1})--(\ref{eq5.22}). In so doing, two  Darboux functions  $~^{cs}E_{21}(n)$ and $~^{cs}E_{23}(n)$  are  proved to be determined explicitly by the \textit{a priori} known  seed values $~^{s}G_{21}(n)$ and $~^{s}G_{23}(n)$ of the prototype functions $G_{21}(n)$ and $G_{23}(n)$.
By  dint of all these observations  the above written formula (\ref{eq4.13}) produces the set of six nonuniform linear equations serving to restore the  unknown Darboux functions  relying upon the known ones. In so doing, the seed solution ~$^{s}X(n|z)$ to the auxiliary linear problem (\ref{eq2.4}) and (\ref{eq2.5}) must be found beforehand.

Once the necessary elements of Darboux matrix have been found, the proper equations taken among  the extended form (\ref{eq5.1})--(\ref{eq5.22}) of implicit B\"{a}cklund transformation (\ref{eq4.2}) allow to obtain explicit crop solutions for the prototype field functions.

\section{Peculiarities of the  Darboux--B\"{a}cklund dressing technique stipulated by the chosen two-fold Darboux matrix}
\label{sec5 }
\setcounter{equation}{0}

As we have already mentioned,   the first (\ref{eq4.2}) of two matrix equations for the Darboux matrix should be treated as the implicit B\"{a}cklund transformation  between the seed  $~^{s}p_{11}(n)$,    $~^{s}p_{13}(n)$, $~^{s}p_{31}(n)$, $~^{s}p_{33}(n)$, $~^{s}F_{12}(n)$, $~^{s}G_{21}(n)$, $~^{s}G_{23}(n)$, $~^{s}F_{32}(n)$ and crop  $~^{c}p_{11}(n)$, $~^{c}p_{13}(n)$, $~^{c}p_{31}(n)$, $~^{c}p_{33}(n)$, $~^{c}F_{12}(n)$, $~^{c}G_{21}(n)$, $~^{c}G_{23}(n)$, $~^{c}F_{32}(n)$  solutions for the prototype  fields. To justify this statement in the case of adopted two-fold ansatz (\ref{eq4.8}) for the Darboux matrix it is sufficient to observe that  the relevant matrix formula (\ref{eq4.2}) comprises the following twenty two equations
\begin{equation}\label{eq5.1}
^{cs}K_{11}(n+1)+~^{cs}C_{12}(n+1)~^{s}G_{21}(n)=
~^{cs}K_{11}(n)+~^{c}F_{12}(n)~^{cs}E_{21}(n)
\end{equation}
\begin{eqnarray}\label{eq5.2}
^{cs}K_{11}(n+1)~^{s}p_{11}(n)&+&~^{cs}T_{12}(n+1)~^{s}G_{21}(n)+~^{cs}K_{13}(n+1)~^{s}p_{31}(n)=
\nonumber\\
&=&^{c}p_{11}(n)~^{cs}K_{11}(n)+~^{c}F_{12}(n)~^{cs}V_{21}(n)+~^{c}p_{13}(n)^{cs}K_{31}(n)
\end{eqnarray}
\begin{equation}\label{eq5.3}
^{cs}C_{12}(n)+~^{c}F_{12}(n)=0
\end{equation}
\begin{equation}\label{eq5.4}
^{cs}T_{12}(n)+~^{c}p_{11}(n)~^{cs}C_{12}(n)+
~^{c}F_{12}(n)~^{cs}D_{22}(n)+~^{c}p_{13}(n)~^{cs}C_{32}(n)=0
\end{equation}
\begin{eqnarray}\label{eq5.5}
^{cs}K_{11}(n+1)~^{s}F_{12}(n)&+&~^{cs}K_{13}(n+1)~^{s}F_{32}(n)=
\nonumber\\
&=&^{c}p_{11}(n)~^{cs}T_{12}(n)+~^{c}F_{12}(n)~^{cs}U_{22}(n)+~^{c}p_{13}(n)^{cs}T_{32}(n)
\end{eqnarray}
\begin{equation}\label{eq5.6}
^{cs}C_{12}(n+1)~^{s}G_{23}(n)+~^{cs}K_{13}(n+1)=
~^{cs}K_{13}(n)+~^{c}F_{12}(n)~^{cs}E_{23}(n)
\end{equation}
\begin{eqnarray}\label{eq5.7}
^{cs}K_{11}(n+1)~^{s}p_{13}(n)&+&~^{cs}T_{12}(n+1)~^{s}G_{23}(n)+~^{cs}K_{13}(n+1)~^{s}p_{33}(n)=
\nonumber\\
&=&^{c}p_{11}(n)~^{cs}K_{13}(n)+~^{c}F_{12}(n)~^{cs}V_{23}(n)+~^{c}p_{13}(n)^{cs}K_{33}(n)
\end{eqnarray}
\begin{equation}\label{eq5.8}
^{cs}E_{21}(n+1)+~^{s}G_{21}(n)=0
\end{equation}
\begin{equation}\label{eq5.9}
^{cs}E_{21}(n+1)~^{s}p_{11}(n)+~^{cs}V_{21}(n+1)+
~^{cs}D_{22}(n+1)~^{s}G_{21}(n)+~^{cs}E_{21}(n+1)~^{s}p_{31}(n)=0
\end{equation}
\begin{eqnarray}\label{eq5.10}
^{cs}V_{21}(n+1)~^{s}p_{11}(n)&+&~^{cs}U_{22}(n+1)~^{s}G_{21}(n)+^{cs}V_{23}(n+1)~^{s}p_{31}(n)=
\nonumber\\
&=&^{c}G_{21}(n)~^{cs}K_{11}(n)+~^{c}G_{23}(n)~^{cs}K_{31}(n)
\end{eqnarray}
\begin{equation}\label{eq5.11}
^{cs}E_{21}(n+1)~^{s}F_{12}(n)+~^{cs}E_{23}(n+1)~^{s}F_{32}(n)=
~^{c}G_{21}(n)~^{cs}C_{12}(n)+~^{c}G_{23}(n)~^{cs}C_{32}(n)
\end{equation}
\begin{equation}\label{eq5.12}
^{cs}V_{21}(n+1)~^{s}F_{12}(n)+~^{cs}V_{23}(n+1)~^{s}F_{32}(n)=
~^{c}G_{21}(n)~^{cs}T_{12}(n)+~^{c}G_{23}(n)~^{cs}T_{32}(n)
\end{equation}
\begin{equation}\label{eq5.13}
^{cs}E_{23}(n+1)+~^{s}G_{23}(n)=0
\end{equation}
\begin{equation}\label{eq5.14}
^{cs}E_{21}(n+1)~^{s}p_{13}(n)+~^{cs}D_{22}(n+1)~^{s}G_{23}(n)+
~^{cs}E_{23}(n+1)~^{s}p_{33}(n)+~^{cs}V_{23}(n+1)=0
\end{equation}
\begin{eqnarray}\label{eq5.15}
^{cs}V_{21}(n+1)~^{s}p_{13}(n)&+&~^{cs}U_{22}(n+1)~^{s}G_{23}(n)+^{cs}V_{23}(n+1)~^{s}p_{33}(n)=
\nonumber\\
&=&^{c}G_{21}(n)~^{cs}K_{13}(n)+~^{c}G_{23}(n)~^{cs}K_{33}(n)
\end{eqnarray}
\begin{equation}\label{eq5.16}
^{cs}K_{31}(n+1)+~^{cs}C_{32}(n+1)~^{s}G_{21}(n)=
~^{c}F_{32}(n)~^{cs}E_{21}(n)+~^{cs}K_{31}(n)
\end{equation}
\begin{eqnarray}\label{eq5.17}
^{cs}K_{31}(n+1)~^{s}p_{11}(n)&+&~^{cs}T_{32}(n+1)~^{s}G_{21}(n)+~^{cs}K_{33}(n+1)~^{s}p_{31}(n)=
\nonumber\\
&=&^{c}p_{31}(n)~^{cs}K_{11}(n)+~^{c}F_{32}(n)~^{cs}V_{21}(n)+~^{c}p_{33}(n)^{cs}K_{31}(n)
\end{eqnarray}
\begin{equation}\label{eq5.18}
^{cs}C_{32}(n)+~^{c}F_{32}(n)=0
\end{equation}
\begin{equation}\label{eq5.19}
^{c}p_{31}(n)~^{cs}C_{12}(n)+~^{c}F_{32}(n)~^{cs}D_{22}(n)=
~^{cs}T_{32}(n)+~^{c}p_{33}(n)~^{cs}C_{32}(n)
\end{equation}
\begin{eqnarray}\label{eq5.20}
^{cs}K_{31}(n+1)~^{s}F_{12}(n)&+&~^{cs}K_{33}(n+1)~^{s}F_{32}(n)=
\nonumber\\
&=&^{c}p_{31}(n)~^{cs}T_{12}(n)+~^{c}F_{32}(n)~^{cs}U_{22}(n)+~^{c}p_{33}(n)^{cs}T_{32}(n)
\end{eqnarray}
\begin{equation}\label{eq5.21}
^{cs}C_{32}(n+1)~^{s}G_{23}(n)+~^{cs}K_{33}(n+1)=
~^{c}F_{32}(n)~^{cs}E_{23}(n)+~^{cs}K_{33}(n)
\end{equation}
\begin{eqnarray}\label{eq5.22}
^{cs}K_{31}(n+1)~^{s}p_{13}(n)&+&~^{cs}T_{32}(n+1)~^{s}G_{23}(n)+~^{cs}K_{33}(n+1)~^{s}p_{33}(n)=
\nonumber\\
&=&^{c}p_{31}(n)~^{cs}K_{13}(n)+~^{c}F_{32}(n)~^{cs}V_{23}(n)+~^{c}p_{33}(n)^{cs}K_{33}(n)\,.
\end{eqnarray}
Provided all fourteen Darboux functions $~^{cs}K_{11}(n)$, $~^{cs}C_{12}(n)$, $~^{cs}T_{12}(n)$, $~^{cs}K_{13}(n)$, $~^{cs}E_{21}(n)$, $~^{cs}V_{21}(n)$,  $~^{cs}D_{22}(n)$, $~^{cs}U_{22}(n)$, $~^{cs}E_{23}(n)$, $~^{cs}V_{23}(n)$, $~^{cs}K_{31}(n)$, $~^{cs}C_{32}(n)$, $~^{cs}T_{32}(n)$, $~^{cs}K_{33}(n)$ in these  detailed relationships (\ref{eq5.1})--(\ref{eq5.22}) have been excluded, we come to just eight equations establishing explicit B\"{a}cklund transformation between eight seed and eight crop prototype field functions.

Though having been very impractical for the actual calculations of solitonic or any other solutions  the   explicit B\"{a}cklund transformation as such provides  a strong argument for the  original implicit  B\"{a}cklund transformation (\ref{eq5.1})--(\ref{eq5.22}) to be naturally incorporated into the  Darboux--B\"{a}cklund integration procedure as its crucial and intrinsically noncontradictory element.

First of all, two    B\"{a}cklund  relations (\ref{eq5.8}) and (\ref{eq5.13}) clearly indicate that two Darboux functions $~^{cs}E_{21}(n)$ and $~^{cs}E_{23}(n)$ are actually known $~^{cs}E_{21}(n)=-~^{s}G_{21}(n-1)$, $~^{cs}E_{23}(n)=-~^{s}G_{23}(n-1)$.
As a consequence, the formulas (\ref{eq5.1}), (\ref{eq5.6}), (\ref{eq5.16}), (\ref{eq5.21}) accompanied by another  two    B\"{a}cklund  relations (\ref{eq5.3}) and (\ref{eq5.18}) yield
\begin{equation}\label{eq5.23}
^{cs}K_{11}(n)=~^{cs}K_{11}-~^{cs}C_{12}(n)~^{s}G_{21}(n-1)
\end{equation}
\begin{equation}\label{eq5.24}
^{cs}K_{13}(n)=~^{cs}K_{13}-~^{cs}C_{12}(n)~^{s}G_{23}(n-1)
\end{equation}
\begin{equation}\label{eq5.25}
^{cs}K_{31}(n)=~^{cs}K_{31}-~^{cs}C_{32}(n)~^{s}G_{21}(n-1)
\end{equation}
\begin{equation}\label{eq5.26}
^{cs}K_{33}(n)=~^{cs}K_{33}-~^{cs}C_{32}(n)~^{s}G_{23}(n-1)\,.
\end{equation}
Here  the coordinate-independent parameters $~^{cs}K_{jk}$ are proved to be time-independent too. To confirm their time-independence it is sufficient to combine the expressions  (\ref{eq5.23})--(\ref{eq5.26}) for $~^{cs}K_{jk}(n)$ with the proper evolution equations selected from the matrix-valued  evolution equation  for the Darboux matrix (\ref{eq4.3}) and to use the equalities  (\ref{eq5.3}), (\ref{eq5.8}), (\ref{eq5.13}), (\ref{eq5.18}).
On the other hand, the formulas (\ref{eq5.9}), (\ref{eq5.14}) yield
\begin{equation}\label{eq5.27}
^{cs}V_{21}(n)=~^{s}G_{21}(n-1)~^{s}p_{11}(n-1)+~^{s}G_{23}(n-1)~^{s}p_{31}(n-1)-~^{cs}D_{22}(n)~^{s}G_{21}(n-1)
\end{equation}
\begin{equation}\label{eq5.28}
^{cs}V_{23}(n)=~^{s}G_{21}(n-1)~^{s}p_{13}(n-1)+~^{s}G_{23}(n-1)~^{s}p_{33}(n-1)-~^{cs}D_{22}(n)~^{s}G_{23}(n-1)\,.
\end{equation}
These  six relationships (\ref{eq5.23})--(\ref{eq5.28}) accompanied by equalities (\ref{eq5.8}) and (\ref{eq5.13}) say that  only six independent Darboux functions $~^{cs}C_{12}(n)$, $~^{cs}T_{12}(n)$, $~^{cs}D_{22}(n)$, $~^{cs}U_{22}(n)$, $~^{cs}C_{32}(n)$, $~^{cs}T_{32}(n)$   have to be found relying upon the spectral properties of Darboux matrix.

By another words, the number of nonuniform linear equations encoded in expression (\ref{eq4.13}) indeed  coincides with the number of unknown independent Darboux functions. Hence, all six independent Darboux functions $~^{cs}C_{12}(n)$, $~^{cs}T_{12}(n)$, $~^{cs}D_{22}(n)$, $~^{cs}U_{22}(n)$, $~^{cs}C_{32}(n)$, $~^{cs}T_{32}(n)$ can be  calculated in terms of seed solution $~^{s}X(n|z)$ to  the  auxiliary linear problem (\ref{eq2.4}) and  (\ref{eq2.5}) supplemented by the spectral characteristics $~^{cs}z_{\pm}$ and $~^{cs}\varepsilon(\,^{cs}z_{\pm})$ of requested Darboux matrix (\ref{eq4.8}).  In so doing, the relevant set  of  six nonuniform linear equations  is split into three separate subsets given by formulas
\begin{eqnarray}\label{eq5.29}
&~&^{cs}C_{12}(n)
\left[~^{s}G_{21}(n-1)~^{s}Y_{1}(n|\,^{cs}z_{\pm})-
\lambda(\,^{cs}z_{\pm})~^{s}Y_{2}(n|\,^{cs}z_{\pm})+
~^{s}G_{23}(n-1)~^{s}Y_{3}(n|\,^{cs}z_{\pm})\,\right]-
\nonumber\\
&-&~^{cs}T_{12}(n)~^{s}Y_{2}(n|\,^{cs}z_{\pm})=
~^{cs}K_{11}~^{s}Y_{1}(n|\,^{cs}z_{\pm})+
~^{cs}K_{13}~^{s}Y_{3}(n|\,^{cs}z_{\pm})
\end{eqnarray}
\begin{eqnarray}\label{eq5.30}
&~&^{cs}D_{22}(n)
\left[~^{s}G_{21}(n-1)~^{s}Y_{1}(n|\,^{cs}z_{\pm})-
\lambda(\,^{cs}z_{\pm})~^{s}Y_{2}(n|\,^{cs}z_{\pm})+
~^{s}G_{23}(n-1)~^{s}Y_{3}(n|\,^{cs}z_{\pm})\,\right]-
\nonumber\\
&-&~^{cs}U_{22}(n)~^{s}Y_{2}(n|\,^{cs}z_{\pm})=
\nonumber\\
&=&\left[~^{s}G_{21}(n-1)~^{s}p_{11}(n-1)+~^{s}G_{23}(n-1)~^{s}p_{31}(n-1)-~^{s}G_{21}(n-1)\,\lambda(\,^{cs}z_{\pm})\,\right]
~^{s}Y_{1}(n|\,^{cs}z_{\pm})+
\nonumber\\
&+&\lambda^{2}(\,^{cs}z_{\pm})~^{s}Y_{2}(n|\,^{cs}z_{\pm})+
\nonumber\\
&+&\left[~^{s}G_{21}(n-1)~^{s}p_{13}(n-1)+~^{s}G_{23}(n-1)~^{s}p_{33}(n-1)-~^{s}G_{23}(n-1)\,\lambda(\,^{cs}z_{\pm})\,\right]
~^{s}Y_{3}(n|\,^{cs}z_{\pm})
\end{eqnarray}
\begin{eqnarray}\label{eq5.31}
&~&^{cs}C_{32}(n)
\left[~^{s}G_{21}(n-1)~^{s}Y_{1}(n|\,^{cs}z_{\pm})-
\lambda(\,^{cs}z_{\pm})~^{s}Y_{2}(n|\,^{cs}z_{\pm})+
~^{s}G_{23}(n-1)~^{s}Y_{3}(n|\,^{cs}z_{\pm})\,\right]-
\nonumber\\
&-&~^{cs}T_{32}(n)~^{s}Y_{2}(n|\,^{cs}z_{\pm})=
~^{cs}K_{31}~^{s}Y_{1}(n|\,^{cs}z_{\pm})+
~^{cs}K_{33}~^{s}Y_{3}(n|\,^{cs}z_{\pm})\,.
\end{eqnarray}
Here the shorthand notation
\begin{equation}\label{eq5.32}
^{s}Y_{j}(n|\,^{cs}z_{\pm})=
\sum_{k=1}^{3} ~^{s}X_{jk}(n|\,^{cs}z_{\pm})~^{cs}\varepsilon(\,^{cs}z_{\pm})
\end{equation}
has been introduced. The first set (\ref{eq5.29}) is seen to produce the expressions for $~^{cs}C_{12}(n)$ and $~^{cs}T_{12}(n)$. The second set (\ref{eq5.30}) gives rise to  the expressions for $~^{cs}D_{22}(n)$ and $~^{cs}U_{22}(n)$. At last, the third set (\ref{eq5.31}) allows to obtain the expressions for $~^{cs}C_{32}(n)$ and $~^{cs}T_{32}(n)$.
By dint of relationships (\ref{eq5.23})--(\ref{eq5.28}) between the dependent and independent  Darboux functions the dependent Darboux functions $~^{cs}K_{11}(n)$, $~^{cs}K_{13}(n)$, $~^{cs}K_{31}(n)$, $~^{cs}K_{33}(n)$ and $~^{cs}V_{21}(n)$, $~^{cs}V_{23}(n)$ are  also readily restorable.

Thus, the procedure of  obtaining the expressions for Darboux functions relying upon the spectral properties of Darboux matrix is seen to be complete .

The next step in our task is to select eight  formulas among the implicit B\"{a}cklund transformation   (\ref{eq5.1})--(\ref{eq5.22}) suitable to convert the expressions for Darboux functions into the expressions for crop field functions. Below we  summarize the recipes for  relevant calculations.

(1) The expressions for $~^{c}F_{12}(n)$ and $~^{c}F_{32}(n)$  are given directly by formulas (\ref{eq5.3}) and (\ref{eq5.18}).

(2) The expressions for $~^{c}G_{21}(n)$ and $~^{c}G_{23}(n)$ follow from the set of two linear equations (\ref{eq5.10}) and (\ref{eq5.15}).

(3) The expressions for $~^{c}p_{11}(n)$ and $~^{c}p_{13}(n)$  are obtainable  from the set of two linear equations (\ref{eq5.2}) and (\ref{eq5.7}) with the relationship (\ref{eq5.3}) for $~^{c}F_{12}(n)$  taken into account.

(4) At last, the expressions for $~^{c}p_{31}(n)$ and $~^{c}p_{33}(n)$  are obtainable  from the set of two linear equations (\ref{eq5.17}) and (\ref{eq5.22}) with the relationship (\ref{eq5.18}) for $~^{c}F_{32}(n)$  taken into account.

\section{Solution to the auxiliary linear problem versus the vacuum seed solution to the prototype nonlinear system}
\label{sec6}
\setcounter{equation}{0}

In order to complete the development of  Darboux--B\"{a}cklund dressing scheme we must  rely upon some known   solution  to the prototype nonlinear system (\ref{eq1.1})--(\ref{eq1.8})  as the seed one. In practice the most preferable  seed solution is  chosen to be a trivial (presumably vacuum) solution ensuring the  respective seed solution to the  auxiliary linear problem (\ref{eq2.4}) and (\ref{eq2.5}) being presentable  in an explicit closed  form.

The most suitable   seed solution to the prototype nonlinear system of our interest (\ref{eq1.1})--(\ref{eq1.8}) looks as follows $~^{s}p_{11}(n)=0$,    $~^{s}p_{13}(n)=0$, $~^{s}p_{31}(n)=0$, $~^{s}p_{33}(n)=0$, $~^{s}F_{12}(n)=~^{0}F_{12}$\,, $~^{s}G_{21}(n)=~^{0}G_{21}$\,, $~^{s}G_{23}(n)=~^{0}G_{23}$\,, $~^{s}F_{32}(n)=~^{0}F_{32}$\,,
where  $~^{0}F_{12}$\,, $~^{0}G_{21}$\,, $~^{0}G_{23}$\,, $~^{0}F_{12}$\, are coordinate- and time-independent quantities subjected to the  natural condition
\begin{equation}\label{eq6.1}
~^{0}F_{12}~^{0}G_{21}+~^{0}F_{32}~^{0}G_{23}=-1\,.
\end{equation}
Consequently, the seed spectral matrix $~^{s}L(n|z)$ and the  seed evolution  matrix $~^{s}A(n|z)$ become coordinate- and time-independent $~^{s}L(n|z)=~^{0}L(z)$ and $~^{s}A(n|z)=~^{0}A(z)$. Moreover,  the zero-curvature  equation (\ref{eq2.1}) yields the strict commutativity relation
\begin{equation}\label{eq6.2}
~^{0}A(z)~^{0}L(z)-~^{0}L(z)~^{0}A(z)=0
\end{equation}
and hence the matrices $~^{0}L(z)$ and $~^{0}A(z)$ must possess the same set of eigenfunctions.
However, the eigenvalues $~^{0}\zeta_{k}(z)$ of matrix $~^{0}L(z)$ are proved to be distinct from  the eigenvalues $~^{0}\eta_{k}(z)$ of matrix $~^{0}A(z)$. Precisely we have
\begin{equation}\label{eq6.3}
^{0}\zeta_{1}(z)=z
\end{equation}
\begin{equation}\label{eq6.4}
^{0}\zeta_{2}(z)=\lambda(z)\equiv z+1/z
\end{equation}
\begin{equation}\label{eq6.5}
^{0}\zeta_{3}(z)=1/z
\end{equation}
and
\begin{equation}\label{eq6.6}
^{0}\eta_{1}(z)=1/z
\end{equation}
\begin{equation}\label{eq6.7}
^{0}\eta_{2}(z)=0
\end{equation}
\begin{equation}\label{eq6.8}
^{0}\eta_{3}(z)=z\,.
\end{equation}
As for the   seed solution $~^{s}X(n|z)=~^{0}X(n|z)$ to the auxiliary linear problem (\ref{eq2.4}) and (\ref{eq2.5}) we come to the expression given by matrix
\begin{equation}\label{eq6.9}
^{0}X(n|z) = \left( \begin{array}{ccccc}
-~^{0}F_{12}~z^{n}\exp(\tau/z)&~~~& +\irm~^{0}G_{23}~[z+1/z]^{n}&~~~& -~^{0}F_{12}~z^{-n}\exp(\tau z)\\[2mm]
~z^{n-1}\exp(\tau/z)&~~~& 0 &~~~& ~z^{1-n}\exp(\tau z)\\[2mm]
-~^{0}F_{32}~z^{n}\exp(\tau/z)&~~~& -\irm~^{0}G_{21}~[z+1/z]^{n}&~~~& -~^{0}F_{32}~z^{-n}\exp(\tau z)\\
\end{array}\right).
\end{equation}

\section{Symmetries  between the prototype field functions and their  implications for the Darboux--B\"{a}cklund scheme}
\label{sec7 }
\setcounter{equation}{0}

The symmetries between the dynamical fields of an integrable nonlinear  dynamical system are known to stipulate  the symmetries between the scattering data of relevant auxiliary spectral problem \cite{vakhnenko-JMP-51-103518, vakhnenko-JPSJ-84-014003, vakhnenko-EPJP-133-243}.
The practical role   of these symmetries for the explicit analytical   integration of  nonlinear systems  is also indisputable \cite{vakhnenko-JMP-51-103518, vakhnenko-JPSJ-84-014003, vakhnenko-EPJP-133-243}.

Analyzing the  prototype nonlinear system of our interest (\ref{eq1.1})--(\ref{eq1.8})  one can readily observe that its equations are consistent   with the  symmetries of complex conjugation
\begin{equation}\label{eq7.1}
p_{11}(n)=p_{33}^{\ast}(n)~~~~~~~~~~~~~~~~~~~~~~~~~~~~~~~~~~~~~~~~p_{33}(n)=p_{11}^{\ast}(n)
\end{equation}
\begin{equation}\label{eq7.2}
p_{13}(n)=p_{31}^{\ast}(n)~~~~~~~~~~~~~~~~~~~~~~~~~~~~~~~~~~~~~~~~p_{31}(n)=p_{13}^{\ast}(n)
\end{equation}
\begin{equation}\label{eq7.3}
F_{12}(n)=F_{32}^{\ast}(n)~~~~~~~~~~~~~~~~~~~~~~~~~~~~~~~~~~~~~~~~F_{32}(n)=F_{12}^{\ast}(n)
\end{equation}
\begin{equation}\label{eq7.4}
G_{21}(n)=G_{23}^{\ast}(n)~~~~~~~~~~~~~~~~~~~~~~~~~~~~~~~~~~~~~~~~G_{23}(n)=G_{32}^{\ast}(n)
\end{equation}
between the prototype field functions.
In view of these symmetries (\ref{eq7.1})--(\ref{eq7.4})  we inevitably come to the following symmetry relations
\begin{equation}\label{eq7.5}
\Omega~\left[L(n|1/z^{\ast})\right]^{\ast}\,\Omega = L(n|z)
\end{equation}
\begin{equation}\label{eq7.6}
\Omega~\left[A(n|1/z^{\ast})\right]^{\ast}\,\Omega = A(n|z)
\end{equation}
for the spectral $L(n|z)$ and evolution $A(n|z)$ matrices written in terms of prototype fields. Here the involutory matrix  $\Omega$  is given by formula
\begin{equation}\label{eq7.7}
\Omega =  \left( \begin{array}{ccccc}
0 &~~&  0  &~~& 1\\
0 &~~&  1  &~~& 0\\
1 &~~&  0  &~~& 0\\
\end{array} \right)\,.
\end{equation}
We impose the similar symmetries
\begin{equation}\label{eq7.8}
\Omega~\left[X(n|1/z^{\ast})\right]^{\ast}\,\Omega = X(n|z)
\end{equation}
\begin{equation}\label{eq7.9}
\Omega~\left[\,^{cs}D(n|1/z^{\ast})\right]^{\ast}\,\Omega = ~^{cs}D(n|z)
\end{equation}
to be valid also for the matrix-solution $X(n|z)$ to the auxiliary linear problem (\ref{eq2.4}) and (\ref{eq2.5}) as well as for the Darboux matrix $~^{cs}D(n|z)$ ~(\ref{eq4.8}).

By the way, the expression (\ref{eq6.9}) for the zero-labeled seed matrix $~^{0}X(n|z)$  has been purposely  arranged to be consistent with the just adopted symmetries.

By means of the last two symmetry relations (\ref{eq7.9}) and (\ref{eq7.8}) one can establish the symmetry relations
\begin{equation}\label{eq7.10}
^{cs}z^{\ast}_{+}~^{cs}z_{-} = 1 =~ ^{cs}z^{\ast}_{-}~^{cs}z_{+}
\end{equation}
and
\begin{equation}\label{eq7.11}
{\left[\,^{cs}\varepsilon_{1}(\,^{cs}z_{+})\right]^{\ast}}/{~^{cs}\varepsilon_{3}(\,^{cs}z_{-})}=1=
{\left[\,^{cs}\varepsilon_{3}(\,^{cs}z_{-})\right]^{\ast}}/{~^{cs}\varepsilon_{1}(\,^{cs}z_{+})}
\end{equation}
\begin{equation}\label{eq7.12}
{\left[\,^{cs}\varepsilon_{2}(\,^{cs}z_{+})\right]^{\ast}}/{~^{cs}\varepsilon_{2}(\,^{cs}z_{-})}=1=
{\left[\,^{cs}\varepsilon_{2}(\,^{cs}z_{-})\right]^{\ast}}/{~^{cs}\varepsilon_{2}(\,^{cs}z_{+})}
\end{equation}
\begin{equation}\label{eq7.13}
{\left[\,^{cs}\varepsilon_{3}(\,^{cs}z_{+})\right]^{\ast}}/{~^{cs}\varepsilon_{1}(\,^{cs}z_{-})}=1=
{\left[\,^{cs}\varepsilon_{1}(\,^{cs}z_{-})\right]^{\ast}}/{~^{cs}\varepsilon_{3}(\,^{cs}z_{+})}
\end{equation}
for the spectral data $~^{cs}z_{\pm}$ and $~^{cs}\varepsilon_{k}(\,^{cs}z_{\pm})$. When tackling   these symmetries (\ref{eq7.10})--(\ref{eq7.13})  the expression (\ref{eq4.11}) for $\det\,^{cs}D(n|z)$ and the condition (\ref{eq4.12})  supporting the property  $\det\,^{c}X(n|~^{cs}z_{\pm})=0$  have also to be  invoked.

\section{Prototype crop solution generated by the two-fold Darboux--B\"{a}cklund transformation}
\label{sec8 }
\setcounter{equation}{0}

In this section we consider the practical implementation of two-fold Darboux--B\"{a}cklund integration scheme with regard to  the  prototype integrable nonlinear system (\ref{eq1.1})--(\ref{eq1.8}). The last four equations (\ref{eq1.5})--(\ref{eq1.8})  of this system clearly indicate that the basic calculations can be safely concentrated only on  four prototype functions $F_{12}(n)$, $G_{21}(n)$, $G_{23}(n)$, $F_{32}(n)$.

To implement our task  let us summarize   the relevant steps  to be made in the framework of  two-fold Darboux--B\"{a}cklund integration  scheme already reported in the previous Sections. Thus, three sets (\ref{eq5.29})--(\ref{eq5.31}) of nonuniform linear equations provide us with the solutions for the Darboux functions  $~^{cs}C_{12}(n)$, $~^{cs}T_{12}(n)$, $~^{cs}D_{22}(n)$, $~^{cs}U_{22}(n)$, $~^{cs}C_{32}(n)$, $~^{cs}T_{32}(n)$. By virtue of relationships (\ref{eq5.23})--(\ref{eq5.26}) the  obtained expressions for $~^{cs}C_{12}(n)$ and  $~^{cs}C_{32}(n)$ give rise to the solutions for the Darboux functions $~^{cs}K_{11}(n)$, $~^{cs}K_{13}(n)$, $~^{cs}K_{31}(n)$, $~^{cs}K_{33}(n)$. The expressions for another two Darboux functions  $~^{cs}V_{21}(n)$ and $~^{cs}V_{23}(n)$ follow from the relationships (\ref{eq5.27}) and (\ref{eq5.28}) by the use of already known expression for $~^{cs}D_{22}(n)$.
Once the explicit expressions for the Darboux functions $~^{cs}C_{12}(n)$ and  $~^{cs}C_{32}(n)$  have been  found, the prototype crop functions $~^{c}F_{12}(n)$,  $~^{c}F_{32}(n)$  are given by  two proper formulas (\ref{eq5.3}) and (\ref{eq5.18})  of implicit B\"{a}cklund transformation. Another two  prototype crop functions $~^{c}G_{21}(n)$  and  $~^{c}G_{23}(n)$ are obtainable as the solutions to the set of two  nonuniform linear equations (\ref{eq5.10}) and (\ref{eq5.15})  of implicit B\"{a}cklund transformation inasmuch as all involved Darboux functions $~^{cs}K_{11}(n)$, $~^{cs}K_{13}(n)$, $~^{cs}K_{31}(n)$, $~^{cs}K_{33}(n)$ and $~^{cs}V_{21}(n)$, $~^{cs}U_{22}(n)$, $~^{cs}V_{23}(n)$   have been restored beforehand.

Though just outlined  dressing  procedure is valid to start  with the prototype seed solution $~^{s}p_{11}(n)$,    $~^{s}p_{13}(n)$, $~^{s}p_{31}(n)$, $~^{s}p_{33}(n)$, $~^{s}F_{12}(n)$\,, $~^{s}G_{21}(n)$\,, $~^{s}G_{23}(n)$\,, $~^{s}F_{32}(n)$\, of an arbitrary order ($s=0, 1, 2, 3,...$),  we apply it to  \textit{a priori} known trivial solution $~^{0}p_{11}(n)=0$,    $~^{0}p_{13}(n)=0$, $~^{0}p_{31}(n)=0$, $~^{0}p_{33}(n)=0$, $~^{0}F_{12}(n)=~^{0}F_{12}$\,, $~^{0}G_{21}(n)=~^{0}G_{21}$\,, $~^{0}G_{23}(n)=~^{0}G_{23}$\,, $~^{0}F_{32}(n)=~^{0}F_{32}$\, labeled by index $s=0$.
Then, taking into account the explicit formula (\ref{eq6.9}) for the zeroth seed solution $~^{0}X(n|z)$ to the auxiliary linear problem (\ref{eq2.4}) and (\ref{eq2.5}) we find  the  prototype crop  functions
$~^{2}F_{12}(n)$, $~^{2}F_{32}(n)$, $~^{2}G_{21}(n)$,  $~^{2}G_{23}(n)$ in the form
\begin{eqnarray}\label{eq8.1}
^{2}F_{12}(n)&=&(\,^{20}K_{11}~^{0}F_{12}+~^{20}K_{13}~^{0}F_{32})\times \nonumber\\
&\times&
\frac{^{0}R(n|\,^{20}z_{+})~^{0}R(n-1|\,^{20}z_{-})-~^{0}R(n|\,^{20}z_{-})~^{0}R(n-1|\,^{20}z_{+})}
{^{0}R(n-1|\,^{20}z_{+})~^{0}R(n-2|\,^{20}z_{-})-~^{0}R(n-1|\,^{20}z_{-})~^{0}R(n-2|\,^{20}z_{+})}-
\nonumber\\
&-&\irm \,(\,^{0}G_{23}~^{20}K_{11}-~^{0}G_{21}~^{20}K_{13})\times \nonumber\\
&\times&
\frac{^{0}S(n|\,^{20}z_{+})~^{0}R(n-1|\,^{20}z_{-})-~^{0}S(n|\,^{20}z_{-})~^{0}R(n-1|\,^{20}z_{+})}
{^{0}R(n-1|\,^{20}z_{+})~^{0}R(n-2|\,^{20}z_{-})-~^{0}R(n-1|\,^{20}z_{-})~^{0}R(n-2|\,^{20}z_{+})}
\end{eqnarray}
\begin{eqnarray}\label{eq8.2}
^{2}F_{32}(n)&=&(\,^{20}K_{33}~^{0}F_{32}+~^{20}K_{31}~^{0}F_{12})\times \nonumber\\
&\times&
\frac{^{0}R(n|\,^{20}z_{+})~^{0}R(n-1|\,^{20}z_{-})-~^{0}R(n|\,^{20}z_{-})~^{0}R(n-1|\,^{20}z_{+})}
{^{0}R(n-1|\,^{20}z_{+})~^{0}R(n-2|\,^{20}z_{-})-~^{0}R(n-1|\,^{20}z_{-})~^{0}R(n-2|\,^{20}z_{+})}+
\nonumber\\
&+&\irm \,(\,^{0}G_{21}~^{20}K_{33}-~^{0}G_{23}~^{20}K_{31})\times \nonumber\\
&\times&
\frac{^{0}S(n|\,^{20}z_{+})~^{0}R(n-1|\,^{20}z_{-})-~^{0}S(n|\,^{20}z_{-})~^{0}R(n-1|\,^{20}z_{+})}
{^{0}R(n-1|\,^{20}z_{+})~^{0}R(n-2|\,^{20}z_{-})-~^{0}R(n-1|\,^{20}z_{-})~^{0}R(n-2|\,^{20}z_{+})}
\end{eqnarray}
\begin{eqnarray}\label{eq8.3}
^{2}G_{21}(n)&=&\frac{^{0}G_{21}~^{20}K_{33}-~^{0}G_{23}~^{20}K_{31}}
{^{20}K_{11}~^{20}K_{33}-~^{20}K_{13}~^{20}K_{31}}\times \nonumber\\
&\times&
\frac{^{0}R(n-1|\,^{20}z_{+})~^{0}R(n-2|\,^{20}z_{-})-~^{0}R(n-1|\,^{20}z_{-})~^{0}R(n-2|\,^{20}z_{+})}
{^{0}R(n|\,^{20}z_{+})~^{0}R(n-1|\,^{20}z_{-})-~^{0}R(n|\,^{20}z_{-})~^{0}R(n-1|\,^{20}z_{+})}
\end{eqnarray}
\begin{eqnarray}\label{eq8.4}
^{2}G_{23}(n)&=&\frac{^{0}G_{23}~^{20}K_{11}-~^{0}G_{21}~^{20}K_{13}}
{^{20}K_{33}~^{20}K_{11}-~^{20}K_{31}~^{20}K_{13}}\times \nonumber\\
&\times&
\frac{^{0}R(n-1|\,^{20}z_{+})~^{0}R(n-2|\,^{20}z_{-})-~^{0}R(n-1|\,^{20}z_{-})~^{0}R(n-2|\,^{20}z_{+})}
{^{0}R(n|\,^{20}z_{+})~^{0}R(n-1|\,^{20}z_{-})-~^{0}R(n|\,^{20}z_{-})~^{0}R(n-1|\,^{20}z_{+})}~.
\end{eqnarray}
Here
\begin{equation}\label{eq8.5}
^{0}R(n|\,^{20}z_{\pm})=
(\,^{20}z_{\pm})^{n}~\exp(\tau/\,^{20}z_{\pm})~^{20}\varepsilon_{1}(\,^{20}z_{\pm})+
(1/\,^{20}z_{\pm})^{n}~\exp(\tau ~^{20}z_{\pm})~^{20}\varepsilon_{3}(\,^{20}z_{\pm})
\end{equation}
\begin{equation}\label{eq8.6}
^{0}S(n|\,^{20}z_{\pm})=~^{20}\varepsilon_{2}(\,^{20}z_{\pm})\,
(\,^{20}z_{\pm}+1/\,^{20}z_{\pm})^{n} .
\end{equation}
In what follows we take
\begin{equation}\label{eq8.7}
^{20}K_{11}=1=~^{20}K_{33}
\end{equation}
and
\begin{equation}\label{eq8.8}
^{20}K_{13}=0=~^{20}K_{31}
\end{equation}
without the loss of generality.

By virtue of the symmetry relations (\ref{eq7.10})--(\ref{eq7.13}) it is reasonable to parameterize the spectral data by formulas
\begin{equation}\label{eq8.9}
^{20}z_{\pm}=\exp(\pm\mu+\irm k)
\end{equation}
\begin{equation}\label{eq8.10}
^{20}\varepsilon_{1}(\,^{20}z_{\pm})=\exp(\pm\alpha+\irm \beta)
\end{equation}
\begin{equation}\label{eq8.11}
^{20}\varepsilon_{2}(\,^{20}z_{\pm})=g\exp(\pm\irm \delta)
\end{equation}
\begin{equation}\label{eq8.12}
^{20}\varepsilon_{3}(\,^{20}z_{\pm})=\exp(\mp\alpha-\irm \beta)\,,
\end{equation}
where $k$, $\alpha$,  $\beta$, $g$,  $\delta$ are arbitrary real parameters, and  $\mu$ is the real  positive  parameter.
To suppress systematic divergences of crop field functions at spatial and temporal infinities we are obliged to impose two  restrictions
\begin{equation}\label{eq8.13}
\mu>\ln(\sqrt{2})
\end{equation}
\begin{equation}\label{eq8.14}
k=\sigma\pi/2
\end{equation}
onto the parameters $\mu$ and $k$ with the coefficient $\sigma$  being defined by the equality
\begin{equation}\label{eq8.15}
\sigma^{2}=1\,.
\end{equation}

In order to apply the general results of this Section  to  an actual nonlinear integrable dynamical system we must adopt some proper parametrization of prototype field variables compatible both with   the  natural constraints (\ref{eq3.17}), (\ref{eq3.18}) and the earlier adopted symmetries (\ref{eq7.1})--(\ref{eq7.4}).

\section{Reduced  ternary system  in the framework of Lagrangian and Hamiltonian formulations}
\label{sec9}
\setcounter{equation}{0}

As we have already mentioned  in Section 3, the natural constraints (\ref{eq3.17}) and (\ref{eq3.18}) are empowered  to   reduce  eight prototype field variables to six actual dynamical ones. In so doing, the  dynamics of the reduced system will be governed by three nonlinear Lagrange equations or alternatively by six nonlinear Hamilton equations.
The key idea to introduce appropriate dynamical variables is to invent parametrization formulas converting both of natural constraints (3.2) and (3.3) into identities with the symmetry relations (7.1)--(7.4) taken into account.
A particular realization  of such   parameterizations is not unique \cite{vakhnenko-JPA-36-5405, vakhnenko-UJP-48-653} and here we consider the following one
\begin{equation}\label{eq9.1}
F_{12}(n)=+\exp[+q_{-}(n)]\sqrt{[1+\irm s(n)]/2}
\end{equation}
\begin{equation}\label{eq9.2}
G_{21}(n)=-\exp[-q_{-}(n)]\sqrt{[1+\irm s(n)]/2}
\end{equation}
\begin{equation}\label{eq9.3}
G_{23}(n)=-\exp[-q_{+}(n)]\sqrt{[1-\irm s(n)]/2}
\end{equation}
\begin{equation}\label{eq9.4}
F_{32}(n)=+\exp[+q_{+}(n)]\sqrt{[1-\irm s(n)]/2}
\end{equation}
\begin{equation}\label{eq9.5}
p_{11}(n)=\frac{1}{4}[\,\dot{q}_{-}(n)-\dot{q}_{+}(n)][1+s^{2}(n)]+
\frac{1}{2}\dot{q}_{-}(n)[1+\irm s(n)]
\end{equation}
\begin{equation}\label{eq9.6}
p_{13}(n)=\frac{\exp[\,q_{-}(n)-q_{+}(n)]}{4\sqrt{1+s^{2}(n)}}
\Bigl\{\dot{q}_{+}(n)[1+s^{2}(n)][1+\irm s(n)]+
\dot{q}_{-}(n)[1+s^{2}(n)][1-\irm s(n)]+2\irm s(n)\Bigr\}
\end{equation}
\begin{equation}\label{eq9.7}
p_{31}(n)=\frac{\exp[\,q_{+}(n)-q_{-}(n)]}{4\sqrt{1+s^{2}(n)}}
\Bigl\{\dot{q}_{-}(n)[1+s^{2}(n)][1-\irm s(n)]+
\dot{q}_{+}(n)[1+s^{2}(n)][1+\irm s(n)]-2\irm s(n)\Bigr\}
\end{equation}
\begin{equation}\label{eq9.8}
p_{33}(n)=\frac{1}{4}[\,\dot{q}_{+}(n)-\dot{q}_{-}(n)][1+s^{2}(n)]+
\frac{1}{2}\dot{q}_{+}(n)[1-\irm s(n)]\,.
\end{equation}
These parametrization formulas (\ref{eq9.1})--(\ref{eq9.8}) are seen to  support the complex conjugate symmetries $q_{-}(n)=q_{+}^{\ast}(n)$ and $q_{+}(n)=q_{-}^{\ast}(n)$ between the field variables $q_{-}(n)$ and $q_{+}(n)$ supplemented by the reality  $s(n)=s^{\ast}(n)$ of field variable $s(n)$. Evidently, the same symmetries take place for the time derivatives of field variables.

By dint of the adopted parameterizations (\ref{eq9.1})--(\ref{eq9.8}), the prototype nonlinear system (\ref{eq1.1})--(\ref{eq1.8}) is reduced to an actual nonlinear lattice system  governed by the set of three Lagrange equations
\begin{equation}\label{eq9.9}
\frac{d}{d\tau} [\partial \L/\partial \dot{q}_{+}(n)]=\partial \L/\partial q_{+}(n)
\end{equation}
\begin{equation}\label{eq9.10}
\frac{d}{d\tau} [\partial \L/\partial \dot{s}(n)]=\partial \L/\partial s(n)
\end{equation}
\begin{equation}\label{eq9.11}
\frac{d}{d\tau} [\partial \L/\partial \dot{q}_{-}(n)]=\partial \L/\partial q_{-}(n)
\end{equation}
with the Lagrange function given by expression
\begin{eqnarray}\label{eq9.12}
\L&=& \frac{1}{4} \,\sum_{m=-\infty}^\infty [1-\irm s(m)]\dot{q}^2_+(m) +
\frac{1}{4} \,\sum_{m=-\infty}^\infty [1+\irm s(m)]\dot{q}^2_-(m) + \frac{1}{8} \,\sum_{m=-\infty}^\infty \,[1+s^2(m)][\dot{q}_+(m)-\dot{q}_-(m)]^2\nonumber\\
&+& \frac{1}{4} \,\sum_{m=-\infty}^\infty  \frac{\dot{s}^2(m)}{1+s^2(m)}  - \frac{1}{2} \,\sum_{m=-\infty}^\infty \exp\bigl[+q_+(m+1)-q_+(m)\bigr] \sqrt{[1-\irm s(m+1)][1-\irm s(m)]} \nonumber\\
&-&\frac{1}{2} \, \sum_{m=-\infty}^\infty  \exp\bigl[+q_-(m+1) -q_-(m)\bigr] \sqrt{[1+\irm s(m+1)][1+\irm s(m)]} +\sum_{m=-\infty}^\infty 1\,.
\end{eqnarray}

The  relevant Hamiltonian formulation of reduced nonlinear system in question is based upon
the set of Hamilton equations
\begin{equation}\label{eq9.13}
\dot{q}_{+}(n)=\partial \H/\partial p_{+}(n)~~~~~~~~~~~~~~~~~~~~~~~~~~~~~~~~~~~~~~~~~~
\dot{p}_{+}(n)=-\partial \H/\partial q_{+}(n)
\end{equation}
\begin{equation}\label{eq9.14}
\dot{s}(n)=\partial \H/\partial r(n)~~~~~~~~~~~~~~~~~~~~~~~~~~~~~~~~~~~~~~~~~~~~~~~~~~
\dot{r}(n)=-\partial \H/\partial s(n)
\end{equation}
\begin{equation}\label{eq9.15}
\dot{q}_{-}(n)=\partial \H/\partial p_{-}(n)~~~~~~~~~~~~~~~~~~~~~~~~~~~~~~~~~~~~~~~~~~
\dot{p}_{-}(n)=-\partial \H/\partial q_{-}(n)
\end{equation}
with the Hamilton function given by expression
\begin{eqnarray}\label{eq9.16}
\H &=& \frac{1}{2} \,\sum_{m=-\infty}^\infty \frac{p^{2}_{+}(m)}{1-\irm s(m)} +
\frac{1}{2} \,\sum_{m=-\infty}^\infty \frac{p^{2}_{-}(m)}{1+\irm s(m)}
+ \frac{1}{4} \,\sum_{m=-\infty}^\infty \,[\,p_{+}(m)+p_{-}(m)]^{2}\nonumber\\
&+&\sum_{m=-\infty}^\infty \,[1+s^{2}(m)] r^{2}(m)
+ \frac{1}{2} \,\sum_{m=-\infty}^\infty \exp\bigl[\,q_+(m+1)-q_+(m)\bigr] \sqrt{[1-\irm s(m+1)][1-\irm s(m)]} \nonumber\\
&+&\frac{1}{2} \, \sum_{m=-\infty}^\infty  \exp\bigl[\,q_-(m+1) -q_-(m)\bigr] \sqrt{[1+\irm s(m+1)][1+\irm s(m)]} -\sum_{m=-\infty}^\infty 1~.
\end{eqnarray}
The standard Hamiltonian form (\ref{eq9.13})--(\ref{eq9.15}) of  above dynamical equations points out on    the standard  form of related  Poisson structure.

In either  of its incarnations (\ref{eq9.9})--(\ref{eq9.12}) or (\ref{eq9.13})--(\ref{eq9.16}) the reduced nonlinear dynamical system comprises three dynamical subsystems coupled both in their kinetic and potential parts and it can be referred to as the nonlinear integrable  ternary  system with the combined inter-mode couplings. Two of its  subsystems  are seen to be qualified  as the  subsystems of complex-valued Toda-like type.

One can readily verify that the Lagrange $\L$ ~(\ref{eq9.12}) and Hamilton $\H$  ~(\ref{eq9.16}) functions are related via   the  Legendre transformation
\begin{equation}\label{eq9.17}
\H+\L=
\sum_{m=-\infty}^\infty[p_{+}(m)\dot{q}_{+}(m)+r(m)\dot{s}(m)+p_{-}(m)\dot{q}_{-}(n)]
\nonumber\\
\end{equation}
with the quantities $p_{+}(n)$, $r(n)$, $p_{-}(n)$  being defined by the conventional  relationships
\begin{equation}\label{eq9.18}
 p_{+}(n)=\partial \L/\partial\dot{q}_{+}(n)
\end{equation}
\begin{equation}\label{eq9.19}
r(n)=\partial \L/\partial\dot{s}(n)
\end{equation}
\begin{equation}\label{eq9.20}
 p_{-}(n)=\partial \L/\partial\dot{q}_{-}(n)\,.
\end{equation}

\section{Modulated pulson  solution to the nonlinear integrable ternary system}
\label{sec10}
\setcounter{equation}{0}

According to  the  Lagrange representation (\ref{eq9.9})--(\ref{eq9.12}) the integration of  reduced nonlinear system  can be safely restricted only by calculations of coordinate field functions $q_{+}(n)$, $s(n)$, $q_{-}(n)$.
To accomplish this purpose we must  invert earlier obtained results (\ref{eq8.1})--(\ref{eq8.4}) for $~^{2}F_{12}(n)$, $~^{2}G_{21}(n)$, $~^{2}G_{23}(n)$, $~^{2}F_{32}(n)$ into formulas for  $~^{2}q_{+}(n)$, $~^{2}s(n)$, $~^{2}q_{-}(n)$  by means of respective  parametrization formulas (\ref{eq9.1})--(\ref{eq9.4}). In so doing,   the specification  formulas (\ref{eq8.5})--(\ref{eq8.15}) supplemented by the parametrization formulas
\begin{equation}\label{eq10.1}
^{0}F_{12}=+\exp(+q_{-})\sqrt{(1+\irm s)/2}
\end{equation}
\begin{equation}\label{eq10.2}
^{0}G_{21}=-\exp(-q_{-})\sqrt{(1+\irm s)/2}
\end{equation}
\begin{equation}\label{eq10.3}
^{0}G_{23}=-\exp(-q_{+})\sqrt{(1-\irm s)/2}
\end{equation}
\begin{equation}\label{eq10.4}
^{0}F_{32}=+\exp(+q_{+})\sqrt{(1-\irm s)/2}
\end{equation}
for the   prototype seed  quantities $~^{0}F_{12}$\,, $~^{0}G_{21}$\,, $~^{0}G_{23}$\,, $~^{0}F_{32}$\, turn out to be useful.
The final result for the crop solution to the reduced integrable nonlinear lattice system of our interest (\ref{eq9.9})--(\ref{eq9.12}) obtainable  in the framework of two-fold Darboux--B\"{a}cklund dressing technique reads as follows
\begin{eqnarray}\label{eq10.5}
q_{+}(n)&=&q_{+}+\frac{1}{2}\ln\left[\frac{\Phi^{2}(n)}{\Phi^{2}(n-1)}\right]+\nonumber\\
&+&\frac{1}{4}\ln \left\{\left[1+s~\frac{\exp(-q_{+}-q_{-})}
{\sqrt{1+s^{2}}}~\frac{\Theta(n)}{\Phi(n)}\right]^{2}+
\frac{\exp(-2q_{+}-2q_{-})}{1+s^{2}}~\frac{\Theta^{2}(n)}{\Phi^{2}(n)}\right\}-\nonumber\\
&-&\frac{\irm}{2} \arctan\left[\frac{\exp(-q_{+}-q_{-})~\Theta(n)}{\sqrt{1+s^{2}}~\Phi(n)+s~\exp(-q_{-}-q_{+})~\Theta(n)}\right]
\end{eqnarray}
\begin{equation}\label{eq10.6}
s(n)=s+\exp(-q_{+}-q_{-})\,\sqrt{1+s^{2}}\,~\frac{\Theta(n)}{\Phi(n)}
\end{equation}
\begin{eqnarray}\label{eq10.7}
q_{-}(n)&=&q_{-}+\frac{1}{2}\ln\left[\frac{\Phi^{2}(n)}{\Phi^{2}(n-1)}\right]+\nonumber\\
&+&\frac{1}{4}\ln \left\{\left[1+s~\frac{\exp(-q_{+}-q_{-})}
{\sqrt{1+s^{2}}}~\frac{\Theta(n)}{\Phi(n)}\right]^{2}+
\frac{\exp(-2q_{+}-2q_{-})}{1+s^{2}}~\frac{\Theta^{2}(n)}{\Phi^{2}(n)}\right\}+\nonumber\\
&+&\frac{\irm}{2} \arctan\left[\frac{\exp(-q_{+}-q_{-})~\Theta(n)}{\sqrt{1+s^{2}}~\Phi(n)+s~\exp(-q_{-}-q_{+})~\Theta(n)}\right]\,.
\end{eqnarray}
Here the upper front  indices in  field functions $~^{2}q_{+}(n)$, $~^{2}s(n)$, $~^{2}q_{-}(n)$ have been omitted for the stylistic purposes  and  the shorthand  notations
\begin{eqnarray}\label{eq10.8}
\Phi(n)=2\sigma\sinh(2\mu n-\mu+2\alpha)+
2\sinh(\mu)
\,\sin[\sigma \pi n-\sigma \pi/2+2\beta-2\tau\sigma\cosh(\mu)]
\end{eqnarray}
\begin{eqnarray}\label{eq10.9}
\Theta(n)&=&g\exp(+\mu n-\mu+\alpha)[2\sinh(\mu)]^{n}
\, \sin[\sigma \pi/2+\delta-\beta+\tau\sigma\exp(-\mu)]+\nonumber\\
&+&g\exp(-\mu n+\mu-\alpha)[2\sinh(\mu)]^{n}
~\sin[\sigma \pi n-\sigma \pi/2+\delta+\beta-\tau\sigma\exp(+\mu)]
\end{eqnarray}
for the typical functional combinations have been introduced.
The obtained solution (\ref{eq10.5})--(\ref{eq10.7}) embraces all three coupled  subsystems to be excited.
Considering the expression (\ref{eq10.8}) for the function $\Phi(n)$ appearing in  denominators of all three components $q_{+}(n)$, $s(n)$, $q_{-}(n)$ of our solution  (\ref{eq10.5})--(\ref{eq10.7}), we clearly see that it has distinct signs at opposite spatial infinities. Consequently, at some unfortunate values of parameter $\alpha$ the quantity $\Phi(n)$ as the function of time $\tau$ can change its sign even at some admissible (\textit{i.e.} integer) values of spatial coordinate $n$. At such spatial coordinates all three  components $q_{+}(n)$, $s(n)$, $q_{-}(n)$ as the functions of time $\tau$ are inflicted by the infinite discontinuities emerging periodically in time with the cyclic frequency  equal to $2\cosh(\mu)$.

In order to eliminate such   nonphysical solutions we must impose the additional condition $\alpha=-\mu l$ onto the parameter $\alpha$, where $l$ is an arbitrary integer.
In this adjusted  case we have $\sinh(2\mu n-\mu -2\mu l)\leq-\sinh(\mu)$ at $n\leq l$ and $\sin(2\mu n-\mu -2\mu l)\geq+\sinh(\mu)$ at $n\geq l+1$. Therefore, the modified quantity
\begin{eqnarray}\label{eq10.10}
\Phi(n)=2\sigma\sinh(2\mu n-\mu-2\mu l)+
2\sinh(\mu)
\, \sin[\sigma \pi n-\sigma \pi/2+2\beta-2\tau\sigma\cosh(\mu)]
\end{eqnarray}
taken as the function of time $\tau$ can acquire zero values only in two neighboring  spatial points $n=l$ and $n=l+1$ never changing its sign in either of them.

As a result, the components $q_{+}(n)$ and $q_{-}(n)$ demonstrate  splashes of infinite amplitudes appearing periodically in time with the cyclic frequency equal to $2\cosh(\mu)$ in three neighboring  spatial points $n=l$,  $n=l+1$, $n=l+2$, while the component $s(n)$ demonstrates splashes only in two points $n=l$ and  $n=l+1$. In so doing, all three components remain being continuous functions of time $\tau$.
Evidently, the splashes taking place  in three distinct spatial points $n=l$,  $n=l+1$, $n=l+2$ occur in different instances of time.
Moreover,    in each of three analyzed formulas (\ref{eq10.5})--(\ref{eq10.7})  we observe contributions modulated by  the  function $\Theta(n)$ characterized by two low cyclic frequencies $\exp(+\mu)$ and $\exp(-\mu)$. For this reason the nonlinear wave packet described by the  obtained solution (\ref{eq10.5})--(\ref{eq10.7}) could be referred to as the  modulated pulson. In general, all three cyclic frequencies $2\cosh(\mu)$ and  $\exp(+\mu)$, $\exp(-\mu)$ are seen to be incommensurable. Hence, the spatial pattern of modulated pulson observed at one instant  of time can never be repeated at any other instant.

\section{Conclusion}
\label{sec11 }
\setcounter{equation}{0}

By writing this paper we have tried to attract attention to the broad  investigation of multi-component semi-discrete nonlinear integrable systems associated with  the auxiliary
linear problems of third and more higher orders. In view of  very sophisticated dynamics prospective for the description of various physical phenomena   such systems require the  development of adequate and understandable  methods of their integration. The first fundamental steps in this direction had been made by Caudrey in his famous work \cite{caudrey-NHMS-97-221} suggested the most general inverse scattering  approach to the integration of semi-discrete nonlinear integrable systems generated by  the auxiliary
linear problems of arbitrary orders. Yet, up to now his pioneer attempts appear to be  underestimated. Maybe such an attitude is caused by the  tedious analytical calculations inevitably arising at practical implementation of Caudrey's integration scheme \cite{vakhnenko-JPA-36-5405, vakhnenko-UJP-48-653}.

In present paper we have considered the alternative approach of integration as applied to the particular prototype semi-discrete nonlinear integrable system associated with  the relevant auxiliary linear problem of third order. The approach is based upon the two-fold Darboux transformation for the seed auxiliary  matrix function accompanied by the implicit B\"{a}cklund transformation for a supposedly known seed solution to the  prototype semi-discrete nonlinear integrable system itself. The majority of calculations have been made in terms of prototype field functions in order that  their results  to be suitable for any reduced  semi-discrete nonlinear integrable system compatible with the so called natural constraints and the adopted symmetries for the  prototype field functions.

We have also used the generalized version of direct recurrent approach to find the main conservation laws applicable to any integrable system associated with the chosen form of third order auxiliary linear problem. One of the obtained  conservation laws originates the Hamilton function of any reduced semi-discrete nonlinear integrable system comprising three coupled dynamical subsystems. This fact is readily verifiable on the  reduced ternary system explicitly written down in terms of two complex-valued Toda-like subsystems  and one intermediate subsystem. The coupling between the Toda-like subsystems is mediated by the  intermediate subsystem both in the kinetic and potential parts of Hamilton function.

Having taken into account the general findings of two-fold Darboux--B\"{a}cklund dressing scheme we have obtained the explicit solution to the reduced ternary system in the form of modulated pulson wave packet.

As the final remark, it is necessary to stress that the complex-valued dynamical fields in the reduced ternary system can be rearranged into the purely real fields responsible for the translational and orientational  modes  typical of long macromolecules both synthesized and natural origins.

\section*{Acknowledgement}
The work has been supported by The National Academy of Sciences of Ukraine within the Project  No~0120U100858 (Functional properties of materials prospective for nanotechnologies).


\end{document}